\documentclass[twocolumn,showpacs]{revtex4}
\usepackage{graphics}
\usepackage{amsmath}
\usepackage{epsf}
\begin{document}

\title{Sympathetic cooling and collisional properties of a Rb-Cs mixture}
\author{M.~Anderlini, E.~Courtade, M. Cristiani, D. Cossart, D. Ciampini, C.
Sias, O.~Morsch, and E.~Arimondo} \affiliation{INFM, Dipartimento
di Fisica E. Fermi, Universit\`{a} di Pisa, Largo B. Pontecorvo 3,
I-56127 Pisa, Italy}

\date{\today}
\begin{abstract}
We report on measurements of the collisional properties of a
mixture of $^{133}$Cs and $^{87}$Rb atoms in a magnetic trap at
$\mu\mathrm{K}$ temperatures. By selectively evaporating the Rb
atoms using a radio-frequency field, we achieved sympathetic
cooling of Cs down to a few $\mu\mathrm{K}$. The inter-species
collisional cross-section was determined through rethermalization
measurements, leading to an estimate of $a_s=595\,a_0$ for the
$s$-wave scattering length for Rb in the $|F=2, m_F=2\rangle$ and
Cs in the $|F=4, m_F=4\rangle$ magnetic states. We briefly
speculate on the prospects for reaching Bose-Einstein condensation
of Cs inside a magnetic trap through sympathetic cooling.
\end{abstract}

\maketitle Laser cooling of neutral atoms, combined with
evaporative cooling in conservative (magnetic and optical) traps,
has led to a number of important breakthroughs in atomic physics,
most notably the observation of Bose-Einstein condensation (BEC) in
a dilute gas of alkali atoms in 1995~\cite{ketterle99}. In recent
years, research on ultra-cold atoms has expanded into the realms of
atomic mixtures. Adding a second atomic species has, among other
things, opened up the possibility to sympathetically cool one
atomic species through collisional energy exchange with the other
species~\cite{CollisionsReview}. On the one hand, this has proved
an invaluable tool for reaching the quantum degeneracy regime with
fermionic atoms for which Pauli blocking reduces the evaporative
cooling efficiency at low temperatures. On the other hand,
ultra-cold atomic mixtures are also interesting in their own right.
In particular, the possibility of creating cold heteronuclear
molecules could be an important ingredient in neutral atom quantum
computing due to the expected large permanent dipole moment of such
molecules. Furthermore, ultracold mixtures can lead to interesting
new quantum phases when loaded into an optical
lattice~\cite{lewenstein04}. A large number of mixtures has been
studied in magneto-optical traps (MOTs)~\cite{telles01,mancini04},
and recent experiments by Kerman {\it et
al.}~\cite{kerman04a,kerman04b} have yielded information about the
rovibrational structure of the $^{85}\mathrm{Rb}^{133}\mathrm{Cs}$
molecule. However only a few combinations of ultra-cold atoms have
been experimentally investigated in conservative traps, among them
Li-Cs~\cite{mosk01}, K-Rb~\cite{ferrari02,jin04}, and
Na-Li~\cite{stan04}.

In this work we study the collisional properties of a mixture of
ultra-cold $^{87}\mathrm{Rb}$ and $^{133}\mathrm{Cs}$ atoms in a
magnetic trap. Both Rb and Cs have been used in laser cooling of
atoms for more than fifteen years now and are important as time and
frequency standards. However, not much is known about their
interatomic potentials and collisional properties. While for the
lighter alkalis the interatomic potentials can be calculated
relatively easily, it has not as yet been possible to do the same
for the Rb-Cs potential. Jamieson {\it et al.}~\cite{jamieson03}
calculated the collisional parameters using several similar-looking
choices for the interatomic potentials, and found that even for
small changes in the potential the $s$-wave scattering length
varies by up to two orders of magnitude.

The experimental apparatus used for this work is similar to our
Rb-BEC setup described in detail elsewhere~\cite{muller00}. We use
a double-chamber vacuum system with a 2D collection MOT in the
upper chamber and a six-beam MOT in the lower chamber. In order to
trap and cool both Rb and Cs atoms, the trapping and repumping
light for the two species is superimposed and the same optics
(mirrors, lenses and waveplates) is used to create the beams for
both MOTs. Once the atoms have been transferred into the lower MOT,
after a brief compressed MOT and molasses phase the trapping beams
are switched off and the atoms are optically pumped into the $|F=2,
m_F=2\rangle$ and $|F=4, m_F =4\rangle$ magnetically trappable
states of Rb and Cs, respectively. Immediately after that, the
time-averaged orbiting potential (TOP) magnetic trap is switched
on. Since the two atomic species have different equilibrium
positions in the magnetic trap (due to gravitational sag), the
positions of the MOTs have to be adjusted accordingly in order to
avoid subsequent oscillations in the TOP trap. This is achieved by
tuning the radiation pressure in the MOT using the
(wavelength-selective) quarter-waveplates inserted into the optical
path of the trapping beams.

In order to demonstrate sympathetic cooling, after loading the
atoms into the magnetic trap we performed circle-of-death
evaporative cooling by continuously reducing the strength of the
rotating bias field. This cooling technique is not
species-selective as the circle-of-death is defined by the rotating
zero of field created by the (static) quadrupole and the (rotating)
bias field. Hence, cooling one or the other species separately
(i.e. loading only that species into the MOT) makes little
difference.

\begin{figure}[ht]
\centering\begin{center}\mbox{\epsfxsize 3.0 in
\epsfbox{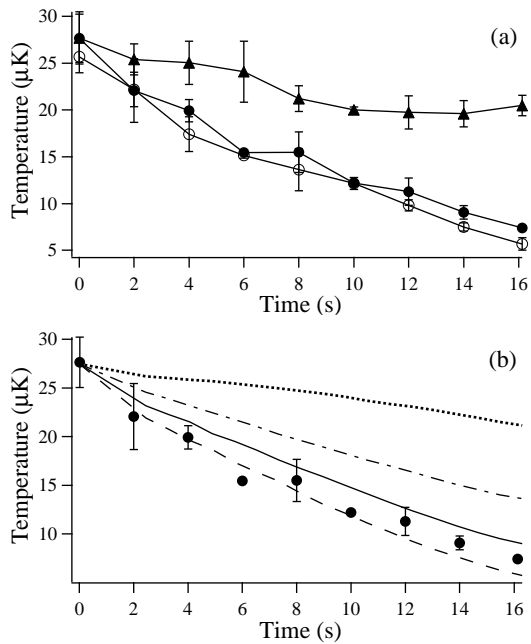}} \caption{(a) Sympathetic cooling of Cs
atoms (filled circles) through RF-evaporation of Rb (open circles).
Removing the Rb atoms before the RF-sweep, Cs atoms are only cooled
slightly by residual circle-of-death evaporation at fixed bias
field (triangles). (b) Comparison of experiment and Monte Carlo
simulations results for Cs: no interspecies collisions (dotted
line), data set C (dash-dotted line), data set A (solid line), data
set B (dashed line). Rb temperatures were only slightly sensitive
of the cross section.}\label{Fig1}
\end{center}
\end{figure}

We then proceeded to apply a radio-frequency field resonant with a
$\Delta m_F = \pm 1$ Zeeman transition in the Rb atoms. By ramping
down the frequency of the RF-field, forced evaporative cooling is
induced as the radius of the surface on which atoms are transferred
into untrapped states shrinks. Radio-frequency evaporation is
species selective as it depends on the Zeeman-sublevel spacing.
Therefore, only the Rb atoms were evaporatively cooled in this way.
Nevertheless, the measured temperature of the Cs atoms exactly
followed the Rb temperature down to a few $\mu \mathrm{K}$, clearly
indicating that sympathetic cooling was taking place (see
Fig.~\ref{Fig1} (a)). In fact, repeating the experiment with Cs
atoms only, a much smaller cooling rate, consistent with residual
circle-of-death evaporation, was observed. Initially, the atom
numbers in the trap were $N_\mathrm{Rb}\approx2\times 10^6$ and
$N_\mathrm{Cs}\approx 10^5$. Towards the end of the ramp, the
number of Rb atoms had dropped by a factor of 10. As expected,
sympathetic cooling was efficient as long as the number of Rb atoms
was larger than the number of Cs atoms; once there were fewer Rb
than Cs atoms, sympathetic cooling stopped.


To understand the dynamics of sympathetic cooling and extract
information on the interspecies scattering properties, we performed
numerical simulations using classical Monte Carlo methods
\cite{AriWw_wuFoot}, extended to samples of two species with
different masses, atom numbers and trap frequencies. Since the
known values of the $C_{6}$ coefficient for the Rb-Cs molecular
potential \cite{marinescu99} give a threshold energy for $p$-wave
scattering of about 50 $\mu K$ \cite{demarco_jin98}, we took into
account both the $s$-wave and the the $p$-wave contributions to the
interspecies elastic cross section. Using the effective range
approximation \cite{joachain,schmidt}, the s-wave term was written
in terms of a scattering length $a$ and an effective range $r_{e}$,
while the $p$-wave term was expressed in terms of a $p$-wave volume
$A_{1}$ \cite{gutierrez}:
\begin{equation}
\sigma_{tot}=\frac{4\pi
a^{2}}{(1-\frac{1}{2}ar_{e}k^{2})^{2}+a^{2}k^{2}}+ \frac{12\pi
A_{1}^{2}k^{4}}{1+A_{1}^{2} k^{6}} \label{TotSigma-Eq}
\end{equation}
where $ k$ is the modulus of the wavevector ${\bf k}=\mu {\bf
v_R}/\hbar$ of the relative motion, with $\mu$ the reduced mass of
the colliding particles. The angular dependence of the $l=1$
partial wave does not affect the efficiency of the p-wave
contribution in the interspecies thermalization process (in
contrast to single-species cross-dimensional relaxation, where it
reduces the $p$-wave contribution by a factor $3/5$
\cite{demarco_jin99}). To analyze the experimental results, we
considered the theoretical sets of scattering parameters
\{$a$,$r_e$,$A_1$\} for the triplet interaction reported in
\cite{jamieson03}, corresponding to data set A = \{595.2$a_0$,
190.2$a_0$, $-168.5\times 10^4\,a_0^3$\}, data set B =
\{177.2$a_0$, 126.4$a_0$, $-4681\times 10^4\,a_0^3$\},  data set C
= \{$-317.6\,a_0$, $424.2\,a_{0}$, $-112.3\times 10^4\,a_0^3$\} and
data set D = \{$-45.37\,a_0$, $3075\,a_{0}$, $-84.22\times
10^4\,a_0^3$\} (where $a_0$ is the Bohr radius). Numerical
simulations using those parameters showed that only the two sets A
and B were consistent with the observed efficiency of the
sympathetic cooling (Fig.~\ref{Fig1} (b)). However, systematic
effects mainly related to the uncertainties in the efficiency of
the evaporation processes prevented an accurate extraction of the
scattering parameters from this data.\\In order to obtain a more
accurate measurement of the scattering cross section, we performed
rethermalization measurements. First, Rb and Cs were cooled down to
temperatures between $5\,\mathrm{\mu K}$ and $60\,\mathrm{\mu K}$
by circle-of-death evaporative cooling. Thereafter, a
radio-frequency ramp was applied in such a way that the Rb atoms
were cooled, but with a fast enough sweep so that there was no
energy exchange between Cs and Rb during the evaporation time. At
the end of the RF ramp there was thus a temperature difference
between the Rb and Cs atoms \cite{Note0}. The mixture was then held
in the magnetic trap for up to $20\,\mathrm{s}$ and the
temperatures of both species were measured as a function of time.
These measurements were repeated for various mean temperatures of
the mixture.

\begin{figure}[ht]
\centering\begin{center}\mbox{\epsfxsize 3.2 in
\epsfbox{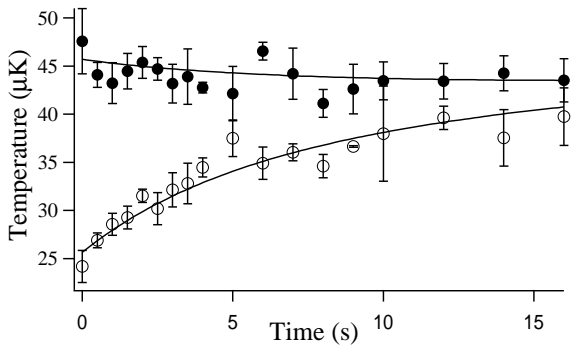}} \caption{Rethermalization in a Rb-Cs
mixture. The initial difference between the Rb temperature (open
circles) and the Cs temperature (filled circles) was created by a
fast RF-evaporation of the Rb atoms. The solid lines are best fits
obtained by a numerical integration of the differential equations
describing the rethermalization process (see text).}\label{Fig2}
\end{center}
\end{figure}

Figure~\ref{Fig2} shows the result of such a rethermalization
measurement. In order to suppress systematic effects as much as
possible, for each experimental run we measured both the Rb and Cs
temperatures {\it without} the other species by eliminating one or
the other through a resonant flash before the rethermalization
process started. Typically, we observed single-species heating
rates of up to $350\,\mathrm{nK\,s^{-1}}$.

Rethermalization techniques have been extensively used to measure
collisional scattering lengths and $p$-wave cross sections
\cite{arndt97,bloch01,ferrari02,schmidt,legere}. For mixtures of
two clouds prepared at different temperatures, the relaxation of
the temperatures due to elastic collisions with a certain constant
cross section proceeds exponentially with a rate that can be
calculated analytically through the model of \cite{mosk01}. From
the rates obtained by exponential fits of observed thermalizations
one can generally extract the value of the cross section. We
extended that model to cross sections that explicitly depend on the
energy of the colliding particles (as in Eq. \ref{TotSigma-Eq}). In
this case the relaxation rate of the temperatures in the mixture
depends linearly on the effective cross section $\sigma_e$
\cite{schmidt,kavoulakis,tolWassen} which, in the case of eq.
\ref{TotSigma-Eq}, is:
\begin{eqnarray}
\sigma_{e}& = &
  \int_{0}^{\infty}dx\,x^5
e^{-x^2}\bigg( \frac{4 \pi a^{2}}{(1-\frac{1}{2}C a r_e x^2)^{2}+ C
a^2 x^2} +\nonumber
\\& & + \frac{ 12 \pi C^2 A_1^{2} x^4}{1+C^3 A_1^2x^6}
\bigg)
\end{eqnarray}
where $C=2\mu k_B(m_1T_2+m_2 T_1)/(\hbar^2 M)$, $M=m_1+m_2$, and
$k_B$ and $\hbar$ are the Boltzmann and the Planck constants,
respectively.

\begin{figure}[ht]
\centering\begin{center}\mbox{\epsfxsize 3.2 in
\epsfbox{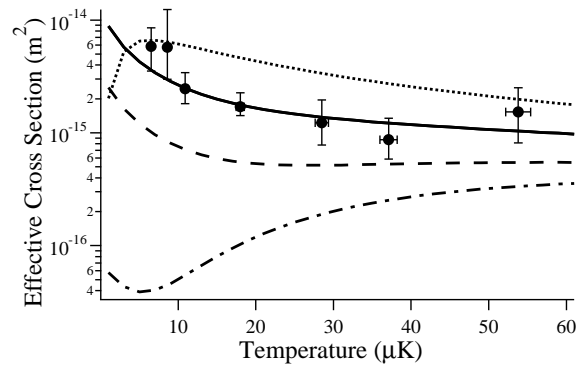}} \caption{Effective elastic scattering cross
sections for Rb-Cs collisions as a function of temperature. The
theoretical predictions by Jamieson {\it et al.}~\cite{jamieson03}
plotted in this graph are: data set B (dotted line), data set A
(solid line), data set C (dashed line) and data set D (dash-dotted
line).}\label{Fig3}
\end{center}
\end{figure}

We checked that exponential fits to this model and to Monte Carlo
simulations gave the same thermalization rates to within a few
percent \cite{Note1}. However, due to the decay of the number of
atoms during the thermalization and to the observed intrinsic
heating independent of the interspecies interactions, we could not
determine the effective cross section through a simple exponential
fit to the data \cite{Note2}. Those cross sections were determined
by running a numerical simulation using the model discussed above
for different $\sigma_{e}$ and initial temperatures $T_0^{Rb}$ and
$T_0^{Cs}$ of the two species, taking into account the experimental
single-species heating rates and atom number decay. We then
compared the results of these simulations with the experimental
data, and from the combination of parameters giving the least
$\chi^2$ we finally determined $\sigma_{e}$. The results of this
analysis are plotted in Fig.~\ref{Fig3} as a function of
temperature (initial weighted average temperature of the mixed
sample).

Having measured the effective scattering cross-sections, we
calculated the effective cross sections corresponding to the
scattering parameters predicted by Jamieson {\it et
al.}~\cite{jamieson03} for the temperature range relevant to our
experiment. The agreement with our data is best for the scattering
parameters of data set A, i.e. for the combination $a=595.2\,a_0$,
$r_e=190.2\,a_0$ and $A_1 =-168.5\times 10^4\, a_0^3$ (see
Fig.~\ref{Fig3}).

 We note here that due to imperfect optical pumping
(and, possibly, other depolarizing processes during the evaporation
cycle), both the Rb and Cs clouds had admixtures of atoms in other
Zeeman sublevels. In the case of Rb, around 90 percent of the atoms
were in the desired state $|F=2, m_F=2\rangle$, with around 10
percent in the $|F=2, m_F=1\rangle$ sublevel, while for Cs we
measured (by performing a Stern-Gerlach type experiment to separate
the Zeeman levels in time-of-flight) relative populations of 70-80
percent in the $|F=4, m_F=4\rangle$ sublevel and 20-30 percent in
$|F=4, m_F=3\rangle$. This posed two problems in interpreting our
data. Firstly, the scattering cross-sections for atoms in the
various Zeeman sublevels are not necessarily the same. Since we
could not eliminate the populations in the other sublevels, we can
only quote them here to indicate the possible error involved in our
determination of the
$\mathrm{Rb}|F=2,m_F=2\rangle-\mathrm{Cs}|F=4,m_F=4\rangle$
cross-section. Secondly, the presence of other Zeeman sublevels
distorted the density profiles from which we calculated the Rb and
Cs temperatures by fitting gaussian curves to the clouds. This
problem was solved by fitting a double-gaussian curve with a fixed
separation (calculated from the known trap parameters and magnetic
moments of the atoms) and extracting the temperature from the
widths of these two gaussians.

Concerning the {\it inelastic} collisional properties of the Rb-Cs
mixture, we found that in the temperature range $5-40\,\mathrm{\mu
K}$ and for typical densities of
$0.5-2\times10^{10}\,\mathrm{cm^{-3}}$, no additional losses in
either Rb or Cs in the presence of the other species occurred. We
can, therefore, put the upper limit of $\sim 10^{-12}
\mathrm{cm^{3}}\mathrm{s^{-1}}$ on the inelastic coefficient for
Rb-Cs collisions in magnetic fields in the range $4-40\,\mathrm{G}$
. In the same range we also performed a slow sweep of the bias
field, but observed no pronounced losses. This rules out the
possibility of broad inter-species Feshbach resonances (as observed
recently for $\mathrm{Na}-\mathrm{Li}$ mixture~\cite{stan04}). In
future experiments it will be interesting to extend the search for
such resonances to higher fields and to other hyperfine states of
Rb and Cs, as they could make possible the creation of ultra-cold
heteronuclear RbCs-molecules directly in the magnetic trap.

Finally, the fact that Cs can be sympathetically cooled by Rb and
that the scattering length for inter-species collisions is large
leads us to speculate whether it might be possible to reach
Bose-Einstein condensation of Cs inside a magnetic trap using a
sympathetic cooling approach. Although Cs was recently condensed
inside an optical trap~\cite{weber03}, it would still be
interesting to achieve condensation in a magnetic trap, thus
avoiding the complicated setup of~\cite{weber03}. Since in our
current experiment the number of Rb atoms we could initially trap
was too small to extend the sympathetic cooling below a few
$\mu\mathrm{K}$, we conducted Monte-Carlo simulations with larger
numbers of atoms. Our simulations took into account the inelastic
losses especially of Cs, which in the $|F=4, m_F=4\rangle$ Zeeman
sublevel has a zero-energy resonance~\cite{soding98} associated
with a large inelastic collision rate responsible for the failure
of all attempts so far to reach Bose-Einstein condensation in a
magnetic trap (for similar reasons, it has not been possible to
condense the $|F=3,m_F=-3\rangle$ level, either). Running
simulations with up to $2\times10^7$ Rb atoms in the $|F=2,
m_F=2\rangle$ sublevel, $10^5$ Cs atoms in $|F=4, m_F=4\rangle$ and
an initial temperature of $10\,\mathrm{\mu K}$ for both species
clearly showed that even with a large (but still realistic) number
of Rb atoms and very low final trap frequencies (chosen so as to
reduce the Cs density and hence inelastic losses), it is not
possible to reach quantum degeneracy in such a scheme. However,
using the $|F=1, m_F=-1\rangle$ and $|F=3,m_F=-3\rangle$ sublevels
for Rb and Cs, respectively, we found that with the same numbers of
atoms as in the first simulation and the same initial temperature,
the threshold for condensation of Cs was reached with roughly
$5\times 10^4$ Cs atoms left in the mixture\cite{note3}.

In conclusion, we have demonstrated sympathetic cooling of Cs atoms
in a Rb-Cs mixture and characterized ultra-cold collisions between
the two species by measuring rethermalization rates. Our results
are consistent with a large interspecies $s$-wave scattering length
around $595\,a_0$. Starting with two orders of magnitude more Rb
atoms (around $2\times 10^7$) than available in the present
experiment, and by trapping Rb and Cs in the lowest respective
hyperfine states, it seems feasible to reach Bose-Einstein
condensation of Cs in a magnetic trap.

We thank T. Bergeman, D. Gu\'{e}ry-Odelin, M. Holland and M.
Jamieson for useful discussions and J.H. M\"{u}ller and N. Malossi
for help in the early stages of the experiment. This research was
supported by the INFM (PRA Photonmatter), by Progetto MIUR-COFIN
2004 and by the EU Network Cold Quantum Gases, contract
HPRN-CT-2000-00125.

\end{document}